\documentclass[aps,prd,twocolumn,nofootinbib]{revtex4-1}
\usepackage{epsfig}
\usepackage[colorlinks,linkcolor=blue,anchorcolor=blue,citecolor=blue,urlcolor=blue,breaklinks=true]{hyperref}
\usepackage{graphics}
\usepackage{slashed}
\usepackage{color}
\usepackage{amsmath}
\usepackage{bm}
\usepackage{setspace}
\usepackage[normalem]{ulem}

\begin{document}
\author{Cheng-Ming Li$^{1}$}
\author{Jin-Li Zhang$^{1}$}
\author{Tong Zhao$^{1}$}
\author{Ya-Peng Zhao$^{1}$}
\author{Hong-Shi Zong$^{1,2,3,}$}\email{zonghs@nju.edu.cn}
\address{$^{1}$ Department of Physics, Nanjing University, Nanjing 210093, China}
\address{$^{2}$Joint Center for Particle, Nuclear Physics and Cosmology, Nanjing 210093, China}
\address{$^{3}$State Key Laboratory of Theoretical Physics, Institute of Theoretical Physics, CAS, Beijing, 100190, China}
\title{Studies of the hybrid star structure within 2+1 flavors NJL model}

\begin{abstract}
In this paper we use the equation of state (EOS) of 2+1 flavors Nambu-Jona-Lasinio (NJL) model to study the structure of compact stars. To avoid the ultraviolet divergence, we employ the proper-time regularization (PTR) with an ultraviolet cutoff. For comparison, we fix three sets of parameters, where the constraints of chemical equilibrium and electric charge neutrality conditions are taken into consideration. With a certain interpolation method in the crossover region, we construct three corresponding hybrid EOSs but find that the maximum masses of hybrid stars in the three different cases don't differ too much. It should be pointed out that, the results we get are in accordance with the recent astro-observation PSR J0348+0432, PSR J1614-2230, PSR J1946+3417.

\bigskip

\noindent Key-words: equation of state, Nambu-Jona-Lasinio model, proper-time regularization, hybrid star
\bigskip

\noindent PACS Numbers: 12.38.Lg, 25.75.Nq, 21.65.Mn

\end{abstract}

\pacs{12.38.Mh, 12.39.-x, 25.75.Nq}

\maketitle

\section{INTRODUCTION}
It's known that the equation of state (EOS) plays a critical role in the study of the structure in compact stars, in which the study of the gravitational mass-radius (M-R) relation is a frontier. If an EOS is given, then combining with the famous Tolman-Oppenheimer-Volkoff (TOV) equations, the M-R relation of a compact star can be derived. Given that the hybrid star is a neutron star with a crust of hadronic matter even hyperonic matter in low density and low temperature, the core is strong interacted quark matter in low temperature and relatively high density, which we have to use Quantum Chromodynamics (QCD) to deal with. However, the perturbation theory is invalid here because of the strong interaction and low energy, thus a nonperturbative theory is needed for this problem. As it is known that the Lattice QCD (LQCD) ~\cite{Borsanyi:2010bpa,PhysRevLett.110.172001} is one of the most reliable approachs to solve the problems of the QCD, however, it doesn't allow to perform calculations at finite baryon chemical potential with three colors because of the "sign problem", therefore we have to choose an effective model to deal with the EOS of the quark core.  There are various alternative effective models for us, such as the Nambu-Jona-Lasinio (NJL) model~\cite{RevModPhys.64.649,Buballa2005205,Cui:2013tva,*Cui:2013aba,NuclPhysB.896.682}, the Dyson-Schwinger equations (DSEs)~\cite{ROBERTS1994477,Roberts2000S1,doi:10.1142/S0218301303001326,Cloet20141,PhysRevD.90.114031,PhysRevD.91.034017,*PhysRevD.91.056003} and the Quantum Electrodynamics in 2+1 dimensions (QED$_3$)~\cite{ROBERTS1994477,PhysRevD.29.2423,PhysRevD.90.036007,PhysRevD.90.073013}. In this paper,we choose the NJL model to study the EOS of the quark matter, since the NJL Lagrangian has many good properties. To be specific, it possesses the feature of dynamical chiral symmetry breaking (DCSB). As we all know, the NJL model can't be renormalized for its four-fermion and six-fermion contact interaction in the Lagrangian. To remedy the shortage, therefore, people usually impose an momentum cutoff on the related integrals, like the three-momentum cutoff regularization. In this work we will adopt the proper-time regularization (PTR)~\cite{NIELSEN1985401,RevModPhys.64.649,Cui:2014hya,doi:10.1142/S0217732316500863,NuclPhysB.896.682} with an ultraviolet (UV) cutoff, which can not only make the momentum integral up to infinity but also avoid the UV divergence with a ``soft'' cutoff. What's more, compared with other regularization frames, PTR also has these features : it's covariant and has an $O(3)$ symmetry for $\mu\neq0$, while for the limit $\mu\rightarrow0$ the $O(4)$ symmetry is restored.

As to the region between the core and the crust of hybrid stars, some authors think that there is a first-order phase transition, while the others argue that there could be a crossover. Recently, the smooth phase transition has been used to study hybrid stars with high mass. But until now, the order of it at zero temperature is still a problem undetermined. In some recent works~\cite{PhysRevC.91.035803,Mallick201496}, the first-order phase transition from Maxwell or Gibbs construction is widely adopted, and the exist of the "mass twins" in the mass-radius relationship for compact stars also seems like to support this viewpoint~\cite{refId0}. However, in astro-observation, it's very difficult to determine the radius of a compact star exactly, thus the "mass twins" can't be find easily. On the other hand, some papers propose that the "mass twins" can also exist in the case of smooth phase transition~\cite{Alvarez-Castillo2015}. In addition, the LQCD shows that the transition line for low net baryon is a crossover~\cite{PhysRevLett.110.172001,PhysRevLett.113.152002,Endrodi:2015oba,Braguta:2015zta}, but model dependent for low temperature and large baryon density. With respect to the phase diagram of the transition, many people consider that there will be a critical end point (CEP), but others do not. Actually, whether there is a CEP and where it is located are still unsolved theoretically. In the Ref.~\cite{Bratovic2013131}, the authors suppose that there is no phase transition but a crossover in the whole phase diagram if the vector interaction is strong enough. What's more, in the Gibbs condition, it's not so reasonable to treat the point-like hadron, which is composed of quarks and gluons, as an independent degree of freedom in the transition region. The fact, thus, supports the study of smooth phase transition in hybrid stars. In recent years, there are some papers studying the massive hybrid stars, which exactly use the smooth crossover as well as different interpolation functions in the phase transition region~\cite{Masuda:2012ed,0004-637X-764-1-12,PhysRevD.91.045003}, while many of these papers have obtained a good result, namely, a phenomenological consequence of the hybrid stars compatible with two solar mass. In this paper, we demonstrate that there are no conspicuous effects on final results among different interpolation methods.

In this paper, we utilize a new EOS from a recent work~\cite{ZhaoYP} which uses the 2+1 flavors NJL model with PTR to study the EOS and obtains a soft EOS for quark matter. After solving the TOV equations, we get the structure of the hybrid stars, which supplies a maximum mass of hybrid stars approximate to 1.9 times of solar mass with the range of radii from 10.8 to 11.2 kilometers. It's significant that the result confirms to the recent measurement and study of PSR J0348+0432, PSR J1614-2230, PSR J1946+3417 ~\cite{Antoniadis1233232,Fonseca:2016tux,doi:10.1093/mnras/stw2947,doi:10.1146/annurev-astro-081915-023322}. On the other hand, the result is insensitive to the choice of the parameters, which illustrates that the model we use is robust in this study. The following of the paper is organized as follows. In Sec.~\ref{one}, we introduce the new EOS of 2+1 flavors NJL model. In Sec.~\ref{two}, a certain interpolation method is used to construct the hybrid EOS which has a smooth phase transition, and for comparison, we apply three sets of parameters to obtain three corresponding hybrid EOSs. Then the M-R relation is deduced in this section. Finally, a brief discussion and summary is given in Sec.~\ref{three}.

\section{The EOS of quark matter with 2+1 flavors NJL model and PTR}\label{one}
In this section, we briefly  introduce the EOS of 2+1 flavors NJL model with  PTR. We can write down the Lagrangian of 2+1 flavors quark system which contains both four-fermion and six-fermion interaction as follows,
\begin{widetext}
\begin{eqnarray}
\mathcal{L}=&&\bar{\psi}i{\not\!\partial}\psi+\sum^8_{i=0}[K_{\rm i}^{(-)}(\bar{\psi}\lambda^i\psi)^2+K_{\rm i}^{(+)}(\bar{\psi}i\gamma^5\lambda^i\psi)^2]+[\frac{1}{2}K_m^{(-)}(\bar{\psi}\lambda^8\psi)(\bar{\psi}\lambda^0\psi)+\frac{1}{2}K_m^{(+)}(\bar{\psi}i\gamma^5\lambda^8\psi)(\bar{\psi}i\gamma^5\lambda^0\psi)]\nonumber\\
&&+[\frac{1}{2}K_m^{(-)}(\bar{\psi}\lambda^0\psi)(\bar{\psi}\lambda^8\psi)+\frac{1}{2}K_m^{(+)}(\bar{\psi}i\gamma^5\lambda^0\psi)(\bar{\psi}i\gamma^5\lambda^8\psi)]+\mathcal{L}_{mass},\,\,\label{effectivelagrangian}
\end{eqnarray}
\end{widetext}
here $K_0^{(\pm)}=G\mp\frac{1}{3}N_{\rm c}K(i\,{\rm tr}S^{\rm s}+2i\,{\rm tr}S^{\rm u})$,

$K_1^{(\pm)}=K_2^{(\pm)}=K_3^{(\pm)}=G\pm\frac{1}{2}N_{\rm c}Ki\,{\rm tr}S^{\rm s}$,

$K_4^{(\pm)}=K_5^{(\pm)}=K_6^{(\pm)}=K_7^{(\pm)}=G\pm\frac{1}{2}N_{\rm c}Ki\,{\rm tr}S^{\rm u}$,

$K_8^{(\pm)}=G\mp\frac{1}{6}N_{\rm c}K(i\,{\rm tr}S^{\rm s}-4i\,{\rm tr}S^{\rm u})$,

$K_{\rm m}^{(\pm)}=G\mp\frac{\sqrt{2}}{3}N_{\rm c}K(i\,{\rm tr}S^{\rm s}-i\,{\rm tr}S^{\rm u})$.

\noindent the trace "tr" is taken in Dirac space, $\lambda^{\rm a}, {\rm a}=1\rightarrow8$ is the Gell-Mann matrix and $\lambda^0$ is defined as $\sqrt{\frac{2}{3}}\,I$ where $I$ is the identity matrix. $G$ and $K$ represent four-fermion and six-fermion interaction coupling constant respectively, $N_{\rm c}=3$ means there are three colors and $\mathcal{L}_{\rm mass}$ is the mass term in the Lagrangian. The $S^{\rm i}$, i=$u$, $d$, $s$ denotes the quark propagator of flavor i, which has a relation to the constituent quark mass $M_{\rm i}$ in the following,
\begin{equation}\label{propagator}
  S_{\rm i}(p^2) = \frac{1}{\not\!p-M_{\rm i}}
\end{equation}
Then we can derive the gap equation,
\begin{eqnarray}
  M_{\rm u} &=& m_{\rm u}-4G\langle\bar{\psi}\psi\rangle_{\rm u}+2K\langle\bar{\psi}\psi\rangle_{\rm u}\langle\bar{\psi}\psi\rangle_{\rm s},\,\,\label{uqge}\\
  M_{\rm s} &=& m_{\rm s}-4G\langle\bar{\psi}\psi\rangle_{\rm s}+2K\langle\bar{\psi}\psi\rangle_{\rm u}^2.\,\,\label{sqge}
\end{eqnarray}
here the symbol $\langle\bar{\psi}\psi\rangle_{\rm u}$ and $\langle\bar{\psi}\psi\rangle_{\rm s}$ represent $u$ and $s$ quark condensate separately (we omit the gap equation of $d$ quark because in our 2+1 flavors NJL model, there is an isospin symmetry between $u$ and $d$ quark, thus their constituent quark mass as well as the quark condensate are equal to each other). According to the definition, the quark condensate can be written as
\begin{eqnarray}
  \langle\bar{\psi}\psi\rangle_{\rm i} &=& -\int\frac{{\rm d}^4p}{(2\pi)^4}{\rm Tr}[iS^{\rm i}(p^2)]\nonumber\\
  &=& -N_{\rm c}\int_{-\infty}^{+\infty}\frac{{\rm d}^4p}{(2\pi)^4}\frac{4iM_{\rm i}}{p^2-M_{\rm i}^2}\,\,\label{qcondensate}
\end{eqnarray}
Here the trace "Tr" is taken in Dirac and color spaces. Until now, the calculations are all done in Minkowski space. However, the non-perturbative theories are often proposed and operated in Euclidean space such as LQCD, because the zero chemical potential Euclidean QCD action defines a probability measure, for which many numerical simulation algorithms are available. Furthermore, working in Euclidean space is more than simply pragmatic: Euclidean lattice field theory is currently a primary candidate for the rigorous definition of an interacting quantum field theory and that relies on it being possible to define the generating functional via proper limiting procedure~\cite{Roberts2000S1}. So we have to translate our calculations from Minkowski space to Euclidean space. And because NJL Lagrangian can't be renormalized, we introduce the PTR, which is defined as
\begin{eqnarray}
  \frac{1}{X^n} &=& \frac{1}{(n-1)!}\int_{0}^{\infty}{\rm d}\tau\tau^{n-1}e^{-\tau X}\nonumber \\
   & &\xrightarrow{\rm UV cutoff} \frac{1}{(n-1)!}\int_{\tau_{\rm UV}}^{\infty}{\rm d}\tau\tau^{n-1}e^{-\tau X},\,\,\label{sregularization}
\end{eqnarray}
Then Eq.~(\ref{qcondensate}) becomes
\begin{eqnarray}
   \langle\bar{\psi}\psi\rangle_i &=& -N_{\rm c}\int_{-\infty}^{+\infty}\frac{{\rm d}^4p^{\rm E}}{(2\pi)^4}\frac{4iM_i}{(p^{\rm E})^{2}+M_i^2}\nonumber\\
   &=& -\frac{N_{\rm c}}{(2\pi)^4}\int_{-\infty}^{+\infty}\int_{-\infty}^{+\infty}{\rm d}^3\overrightarrow{p}{\rm d}p_4\frac{4M_i}{p_4^2+\overrightarrow{p}^2+M_i^2}\nonumber \\
  &=& -\frac{3M_i}{\pi^2}\int_{0}^{+\infty}{\rm d}p\frac{p^2}{\sqrt{p^2+M_i^2}}\nonumber \\
   &=& -\frac{3M_i}{\pi^{\frac{2}{5}}}\int_{\tau_{\rm UV}}^{\infty}\int_{0}^{+\infty}{\rm d}\tau {\rm d}p\tau ^{-\frac{1}{2}}p^2e^{-\tau (M_i^2+p^2)}\nonumber \\
   &=& -\frac{3M_i}{4\pi^2}\int_{\tau_{\rm UV}}^{\infty}{\rm d}\tau \frac{e^{-\tau M_i^2}}{\tau^2}.\,\,\label{regofqcondensate}
\end{eqnarray}
where the superscript E means in Euclidean space.

In order to proceed the following study, we should determine the parameters: two coupling constants, two integral limits, two constituent quark masses (i.e., $M_{\rm u}$ and $M_{\rm s}$, because in SU(2)$_{\rm f}$ symmetry, $M_{\rm u}=M_{\rm d}$, but $M_{\rm s}$ is much larger than $M_{\rm u}$ and $M_{\rm d}$) and one adjustable parameter namely $u$ quark condensate. Specifically, we fit the five of these parameters with five equations and five experimental observable dates, while the other two parameters (the infrared momentum cutoff $\Lambda_{\rm IR}=$235 MeV and the $u$ quark condensate) are fixed before the fitting. According to QCD sum rules and the results of LQCD, the $u$ quark condensate is in the range of -(220 MeV)$^3$$\rightarrow$-(280 MeV)$^3$, thus we choose three point to fix the parameters, namely $\langle\bar{u}u\rangle|_{T=0, \mu=0} =-(230$ MeV$)^3$, $\langle\bar{u}u\rangle|_{T=0, \mu=0} =-(250$ MeV$)^3$ and $\langle\bar{u}u\rangle|_{T=0, \mu=0} =-(270$ MeV$)^3$. On the other hand, we choose $f_{\pi}=$90 MeV, $M_{\pi}=$135 MeV, $M_{K^0}=$500 MeV, $M_{\eta}=$550 MeV, $M_{\eta '}=$950 MeV from the experimental data. Then we fit the parameters to get the following results demonstrated in Table.~\ref{parameters}\footnote{the units of the coupling constants $G$ and $K$ are MeV$^{-2}$ and MeV$^{-5}$ respectively, while the other parameters in this table have the unit of MeV.}.
\begin{widetext}
\begin{center}
\begin{table}[h!]
\caption{Parameter set fixed in our work.}\label{parameters}
\begin{tabular}{p{1.5cm} p{1.3cm} p{2.6cm} p{2.6cm} p{1.3cm}p{1.3cm}p{1.3cm}p{1.3cm}}
\hline\hline
$-(\langle\bar{u}u\rangle)^{\frac{1}{3}}$&$\Lambda_{\rm UV}$&$G$&$K$&$M_{\rm u}$&$M_{\rm s}$&$m_{\rm u}$&$m_{\rm s}$\\
\hline
230&930&$3.48\times10^{-6}$&$1.55\times10^{-13}$&235&455&6&190\\
\hline
250&1080&$2.39\times10^{-6}$&$7.33\times10^{-14}$&215&410&5&150\\
\hline
270&1200&$1.88\times10^{-6}$&$4.45\times10^{-14}$&205&385&4&125\\
\hline\hline
\end{tabular}
\end{table}
\end{center}
\end{widetext}

It's noticed that the previous work is done at zero temperature and zero chemical potential. Next we generalize it into the condition of zero temperature and finite chemical potential, which will make some difference. In Euclidean space, introducing the chemical potential at zero temperature is equivalent to perform a transformation~\cite{PhysRevC.71.015205} that
\begin{equation}\label{muinpfour}
  p_4\rightarrow p_4+i\mu .
\end{equation}
Then we can obtain the analytical result of quark condensate as follows,
\begin{widetext}
\begin{eqnarray}
  \langle\bar{\psi}\psi\rangle_{\rm i} &=& -N_{\rm c}\int_{-\infty}^{+\infty}\frac{{\rm d}^4p}{(2\pi)^4}\frac{4M_{\rm i}}{(p_4+i\mu)^2+M_{\rm i}^2+\overrightarrow{p}^2}\nonumber \\
   &=& -\frac{3M_{\rm i}}{\pi^3}\int_{0}^{+\infty}{\rm d}p\int_{-\infty}^{+\infty}{\rm d}p_4\frac{p^2}{(p_4+i\mu)^2+M_{\rm i}^2+p^2}\nonumber\\
   &=& \left\{
  \begin{array}{lcl}
\displaystyle{-\frac{3M_{\rm i}}{\pi^2}\int_{\sqrt{\mu^2-M_{\rm i}^2}}^{+\infty}{\rm d}p\textstyle{\frac{\left[1-{\rm Erf}(\sqrt{M_{\rm i}^2+p^2}\sqrt{\tau_{\rm UV}})\right]p^2}{\sqrt{M_{\rm i}^2+p^2}}}},\,\,\,\,   M_{\rm i}<\mu\,\,\,\label{mutoqc}\\
\displaystyle{\frac{3M_{\rm i}}{4\pi^2}\left[\textstyle{-M_{\rm i}^2{\rm Ei}(-M_{\rm i}^2\tau_{\rm UV})-\frac{e^{-M_{\rm i}^2\tau_{\rm UV}}}{\tau_{\rm UV}}}\right]},\,\,\,\,\,\,\,\quad\quad M_{\rm i}>\mu
  \end{array}\right.\qquad\qquad
\end{eqnarray}
\end{widetext}
here Ei(x) is an Exponential Integral function which is defined as Ei(x)$=-\int_{-x}^{+\infty}{\rm d}y\frac{e^{-y}}{t}$ and Erf(x) is the error function defined as Erf(x)$=\frac{2}{\sqrt{\pi}}\int_{0}^{x}e^{-\eta^2}{\rm d}\eta$. We can see that the quark condensate is a function of its constituent mass and the chemical potential. Particularly, when $\mu<M_{\rm i}$ the quark condensate will be independent of chemical potential, which is the same as Ref.~\cite{PhysRevD.58.096007}. Actually, if we substitute Eq.~(\ref{mutoqc}) to Eq.~(\ref{uqge}) and~(\ref{sqge}) and solve the iteration equation, we can obtain the constituent quark mass only dependent on the chemical potential.

It's known that the quark number density depends on the chemical potential at zero temperature, which is derived in the following
\begin{eqnarray}
  \rho_{\rm i}(\mu) &=& \langle\psi^+\psi\rangle_{\rm i}\nonumber \\
   &=& -N_{\rm c} \int\frac{{\rm d}^4p}{(2\pi)^4}tr\left[iS_{\rm i}\gamma_0\right]\nonumber\\
   &=& 2N_{\rm c}\int\frac{{\rm d}^3p}{(2\pi)^3}\theta(\mu-\sqrt{p^2+M_{\rm i}^2})\nonumber\\
   &=& \left\{
\begin{array}{lcl}
 \frac{1}{\pi^2}(\sqrt{\mu^2-M_{\rm i}^2})^3,             & &\mu>M_{\rm i}\\
  0,                                                & &\mu<M_{\rm i}
   \end{array}
   \right.\label{qnd}
\end{eqnarray}
\begin{figure}
\includegraphics[width=0.47\textwidth]{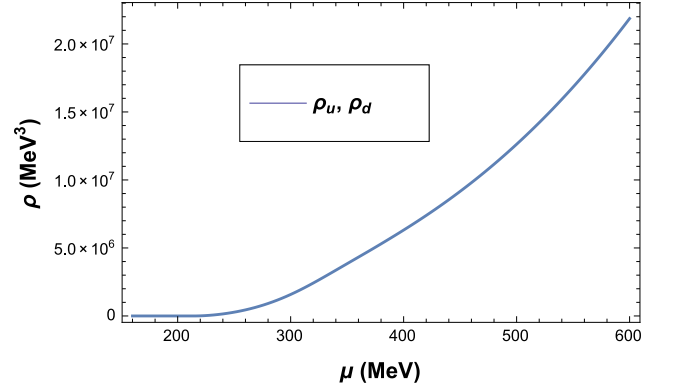}
\caption{quark number density of $u$ and $d$ quark as a function of $\mu$ at $T=0$ with parameters fixed at $\langle\bar{u}u\rangle|_{T=0, \mu=0} =-(230$ MeV$)^3$}
\label{Fig:ndud}
\end{figure}
\begin{figure}
\includegraphics[width=0.47\textwidth]{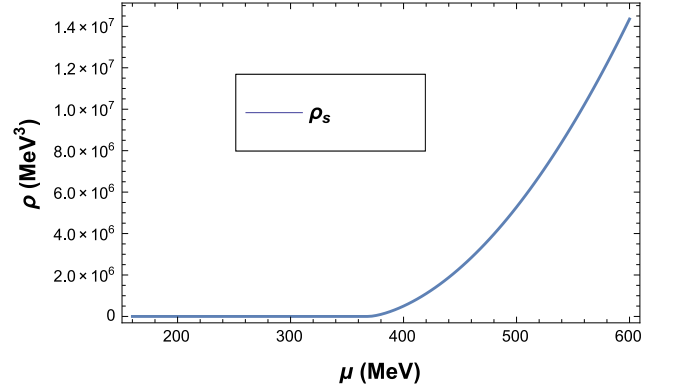}
\caption{quark number density of $s$ quark as a function of $\mu$ at $T=0$ with parameters fixed at $\langle\bar{u}u\rangle|_{T=0, \mu=0} =-(230$ MeV$)^3$}
\label{Fig:nds}
\end{figure}
As we have already got the chemical potential dependance on constituent quark mass, thus we can generalize the same dependance on quark number density, the result of which is shown in Fig.~\ref{Fig:ndud} and Fig.~\ref{Fig:nds} with parameters fixed at $-(\langle\bar{u}u\rangle)^{\frac{1}{3}}=230$ MeV. In these two figures, we can see that for $u$ and $d$ quarks, the critical point is at around $\mu_c=210$ MeV where the quark number density turns to be nonzero with chemical potential increasing; for $s$ quark, $\mu_c=370$ MeV.

If we take the electroweak reactions into account, we have to consider the constraints of chemical equilibrium and electric charge neutrality conditions,
\begin{equation}\label{constrains}
  \left\{\begin{array}{lcl}
           \mu_{\rm d}=\mu_{\rm u}+\mu_{\rm e} \\
           \mu_{\rm s}=\mu_{\rm u}+\mu_{\rm e} \\
           \frac{2}{3}\rho_{\rm u}-\frac{1}{3}\rho_{\rm d}-\frac{1}{3}\rho_{\rm s}-\rho_{\rm e}=0
         \end{array}\right.
\end{equation}
where the particle number density of electron at $T=0$ is $\rho_{\rm e}(\mu_{\rm e})=\frac{\mu_{\rm e}^3}{3\pi^2}$.

Now we can conclude that the three quark number densities are all dependent on the chemical potential of one flavor, which we choose $u$ quark chemical potential in this paper.

By definition, the EOS of QCD at zero temperature and nonzero chemical potential is~\cite{doi:10.1142/S0217751X08040457}
\begin{equation}\label{EOSofQCD}
  P(\mu)=P(\mu=0)+\int_{0}^{\mu}d\mu'\rho(\mu'),
\end{equation}
and the relation between the energy density of the system and its pressure is expressed as~\cite{PhysRevD.86.114028,PhysRevD.51.1989}
\begin{equation}\label{rbedasp}
  \epsilon=-P+\sum_{i}\mu_{\rm i}\rho_{\rm i}
\end{equation}

\section{The structure of hybrid stars with an EOS of smooth phase transition}\label{two}
In convention, if one wants to study the structure of a bare quark star, solving the Tolman-Oppenheimer-Volkoff (TOV) equations with an EOS is necessary,
\begin{eqnarray}
  \frac{{\rm d}p(r)}{{\rm d}r} &=& -\frac{G(\epsilon+P)(M+4\pi r^3P)}{r(r-2GM)} \,\, ,\nonumber\\
  &&\frac{{\rm d}M(r)}{{\rm d}r} = 4\pi r^2\epsilon\,\,\, .\label{TOV}
\end{eqnarray}

In this paper, we choose the APR EOS with $A18+\delta\nu+UIX^{\ast}$ interaction as the EOS of hadronic matter~\cite{PhysRevC.58.1804}, which is proposed using the Argonne $\nu_{18}$ two-nucleon interaction and boost corrections to the two-nucleon interaction as well as three-nucleon interaction. In other words, the system of the APR model is considered as the charge-neutral and beta-stable fluid whose pressure and baryon
chemical potential are equilibrated. Nonetheless, the APR model only considers the degrees of freedom in nucleon system but excludes hyperons because of their unknown interactions. On the whole, the APR model provides a reasonable EOS of hadrons which are only composed of light quarks.

It's noticed that there is still a problem not solved yet, namely, the determination of the term $P(\mu=0)$ in Eq.~(\ref{EOSofQCD}), which is irrelevant to the chemical potential as we can see. Actually, $P(\mu=0)$ represents the pressure of the vacuum, which can't be calculated in a model independent way. Therefore we treat it as a phenomenological parameter standing for the negative pressure of vacuum at zero chemical potential, which manifests the confinement of QCD just as what MIT bag model does. But the ascertaining of the term $P(\mu=0)$ should be legitimate. Like the Ref.~\cite{PhysRevD.92.054012} we identify $P(\mu=0)$ with -$B$ ($B$ is the vacuum bag constant) and choose $B=(120$ MeV$)^4$. In fact, a smaller $B$ can make the EOS stiffer and the upper limit of mass higher. However, to make sure that the energy of quark matter not smaller than hadronic matter in the region of low baryon number density, it's not appropriate to select an excessively small $B$.

The next step is to choose a suitable interpolation function to generate a smooth phase transition between hadronic matter phase and quark matter phase. In the Ref.~\cite{Masuda01072013}, the authors use the P-interpolation and $\epsilon$-interpolation in P-$\rho$ and $\epsilon-\rho$ plane respectively, and in the Ref.~\cite{PhysRevD.92.054012}, the authors also use P-interpolation but in P-$\mu$ plane. In our work, we choose the interpolation method same with the Ref.~\cite{PhysRevD.92.054012}. For detailed, the interpolation function is defined as follows,
\begin{eqnarray}
  P(\mu) &=& P_{\rm H}(\mu)f_-(\mu)+P_{\rm Q}(\mu)f_+(\mu)\,\,,\nonumber\\
  f_{\pm}(\mu) &=& \frac{1}{2}(1\pm {\rm tanh}\,(\frac{\mu-\bar{\mu}}{\Gamma}))\,\,.\label{interpolation}
\end{eqnarray}
\begin{figure}
\includegraphics[width=0.47\textwidth]{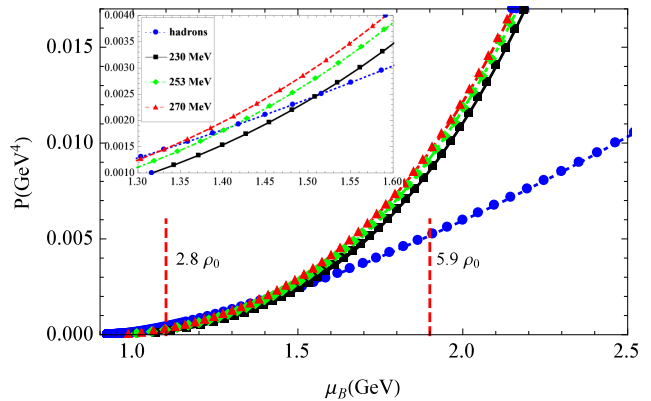}
\caption{the EOS of hadrons which is depicted as the blue dotted line and the EOSs of quark matter which are depicted as the black solid line, the green dotdashed line and the red dashed line corresponding to $\langle\bar{u}u\rangle|_{T=0, \mu=0} =-(230$ MeV$)^3$, $\langle\bar{u}u\rangle|_{T=0, \mu=0} =-(250$ MeV$)^3$, $\langle\bar{u}u\rangle|_{T=0, \mu=0} =-(270$ MeV$)^3$ respectively.}
\label{Fig:twoEOS}
\end{figure}
\begin{figure}
  \includegraphics[width=0.47\textwidth]{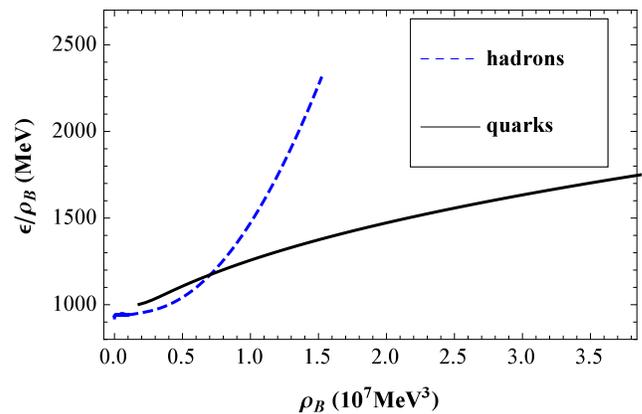}
  \caption{when $\langle\bar{u}u\rangle|_{T=0, \mu=0} =-(230$ MeV$)^3$, the binding energy of quarks and hadrons are shown with the black solid line and the green dashed line respectively.}\label{Fig:bindingenergy}
\end{figure}
here $P_{\rm H}$ and $P_{\rm Q}$ are the pressure in hadronic matter and quark matter respectively. The interpolation functions $f_{\pm}$ are companied with $P_{\rm H}$ and $P_{\rm Q}$ to realize a crossover in the phase transition region. The window $\bar{\mu}-\Gamma\lesssim\mu\lesssim\bar{\mu}+\Gamma$ characterizes the range of the crossover region where both hadrons and quarks are strong interacted, so that neither pure hadronic EOS nor quark EOS are reliable, and in this paper we choose $\Gamma=0.4$ GeV . As we can see in Fig.~\ref{Fig:twoEOS}, when parameters fixed in Table.~\ref{parameters}, the phase transition points are around the intersection of quark EOS curves and hadronic EOS curves, i.e., the baryon chemical potential $\mu_{\rm B}=1.3$ GeV, 1.4 GeV and 1.5 GeV respectively. Thus in this paper we choose the baryon chemical potential from 1.1 GeV to 1.9 GeV as the crossover region. Given the baryon number density of normal nuclear matter $\rho_0=0.17$ fm$^{-3}$, the boundary of the crossover region should correspond to $2.8\rho_0$ and $5.9\rho_0$, which is labeled in Fig.~\ref{Fig:twoEOS}.
\begin{figure}
  \centering
  \includegraphics[width=0.47\textwidth]{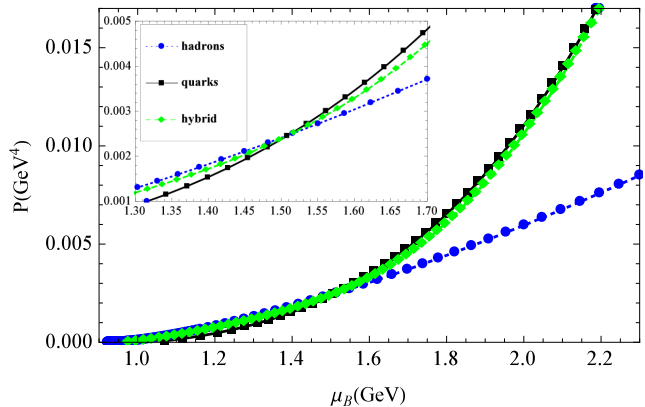}
  \caption{when $\langle\bar{u}u\rangle|_{T=0, \mu=0} =-(230$ MeV$)^3$, the EOS of hybrid system is shown with the green dashed line which is obtained by an interpolation function of hadronic matter EOS and quark matter EOS in the crossover region}\label{Fig:hybridEOS}
\end{figure}
Actually, compared with the binding energy of quarks system and hadrons in APR model which are depicted in Fig.~\ref{Fig:bindingenergy}, when $\rho_{\rm B}<7\times10^6$ MeV$^3$ (namely, 5.15 $\rho_0$), the hadron system is more stable than quarks system. On the contrary, when $\rho_{\rm B}>7\times10^6$ MeV$^3$, the quarks system is more stable. Therefore it's reasonable to employ the interpolation method in this paper.

\begin{figure}
  \centering
  \includegraphics[width=0.47\textwidth]{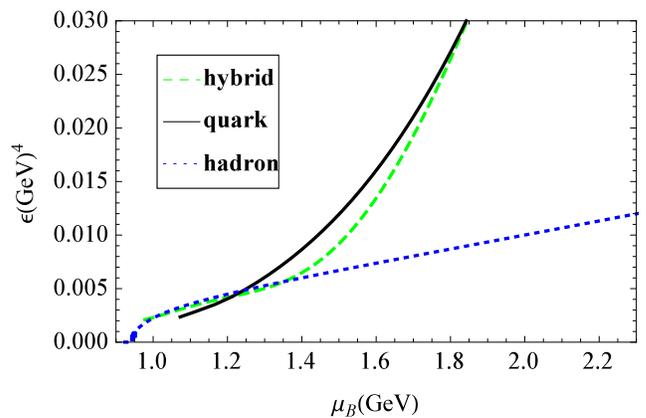}
  \caption{when $\langle\bar{u}u\rangle|_{T=0, \mu=0} =-(230$ MeV$)^3$, the relation between energy density and baryon chemical potential in the hybrid star (blue dotted line), the quark matter (black solid line) and hadronic matter (green dashed line) respectively}\label{Fig:energyandchemicalpotential}
\end{figure}

\begin{figure}
  \centering
  \includegraphics[width=0.47\textwidth]{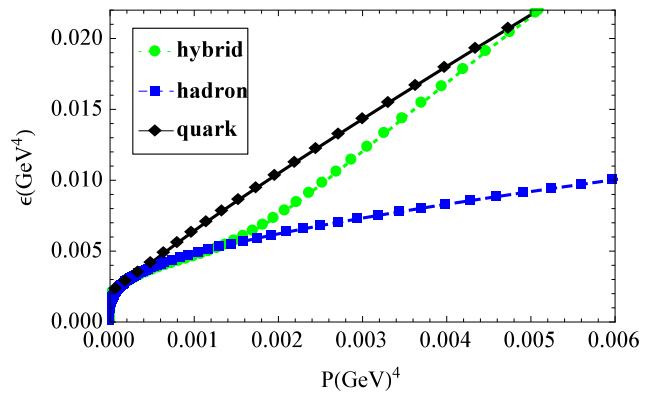}
  \caption{when $\langle\bar{u}u\rangle|_{T=0, \mu=0} =-(230$ MeV$)^3$, the energy density as a function of the pressure in the hybrid star (blue dotted line), quark matter (black solid line) and the hadronic matter (green dashed line)}\label{Fig:energyandpressure}
\end{figure}

With the method mentioned above, we can now deduce the EOS of the whole region, for example, the Fig.~\ref{Fig:hybridEOS} when $\langle\bar{u}u\rangle|_{T=0, \mu=0} =-(230$ MeV$)^3$. It's obvious that when $\mu_{\rm B}<$1.5 GeV, the EOS of the hybrid system (dotted line) is near the EOS of the hadronic matter (dashed line), but when $\mu_{\rm B}>$1.5 GeV it's close to the EOS of quark matter (solid line), the result of which is similar to the other two cases when $\langle\bar{u}u\rangle|_{T=0, \mu=0} =-(250$ MeV$)^3$ and $\langle\bar{u}u\rangle|_{T=0, \mu=0} =-(270$ MeV$)^3$. Furthermore, because the range of the window we choose in this paper is rather small than the corresponding one of the Refs.~\cite{Masuda01072013,PhysRevD.92.054012}, the hybrid EOS will be more reasonable to the real hybrid stars which are usually predicted to have a layer of tens of meters. Through the result obtained in Fig.~\ref{Fig:hybridEOS}, we can also get the relation between energy density and baryon chemical potential as well as the relation between energy density and pressure in Fig.~\ref{Fig:energyandchemicalpotential} and Fig.~\ref{Fig:energyandpressure}.

\begin{figure}
  \centering
  \includegraphics[width=0.47\textwidth]{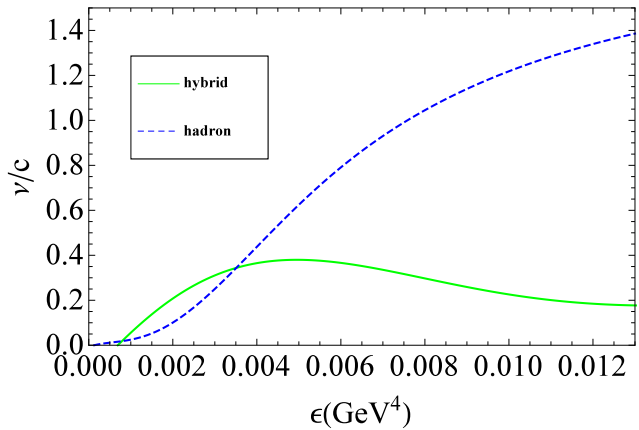}
  \caption{the sound velocity of hadron system in APR model and our hybrid system when $\langle\bar{u}u\rangle|_{T=0, \mu=0} =-(230$ MeV$)^3$ which are depicted as blue dashed line and green solid line respectively.}\label{Fig:soundvelocity}
\end{figure}

In this work, to investigate the rationality of the hybrid EOS, we also calculate the sound velocity of it, which can reflect the stiffness of the system. According to the definition, the sound velocity of a system is
\begin{equation}\label{soundvelocity}
 \nu_{\rm s} = \sqrt{\frac{{\rm d}p}{{\rm d}\epsilon}},
\end{equation}
which is actually concerned with the slope of $P(\epsilon)$ function. In principle, the sound velocity should be smaller than light, and a smaller sound velocity corresponds to a softer EOS. As we can see in Fig.~\ref{Fig:soundvelocity}, the sound velocity of hadron system in APR model is larger than light when the energy density is lager than $8\times10^{-3}$ GeV$^4$, which is not reasonable in reality, while for the hybrid EOS, when $\langle\bar{u}u\rangle|_{T=0, \mu=0} =-(230$ MeV$)^3$, the sound velocity is always smaller than 0.4 times of light velocity.
\begin{figure}
  \centering
  \includegraphics[width=0.47\textwidth]{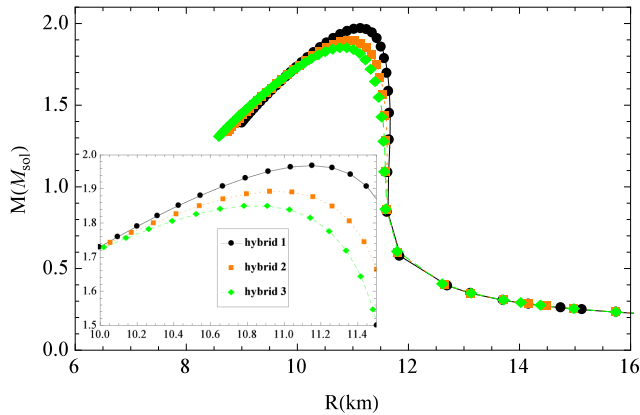}
  \caption{M-R relation of hybrid stars given by three hybrid EOSs of the system, that is, the black solid line corresponds to the case of $\langle\bar{u}u\rangle|_{T=0, \mu=0} =-(230$ MeV$)^3$, the dotted line corresponds to the case of $\langle\bar{u}u\rangle|_{T=0, \mu=0} =-(250$ MeV$)^3$, and the dashed line corresponds to the case of $\langle\bar{u}u\rangle|_{T=0, \mu=0} =-(270$ MeV$)^3$}\label{Fig:massradiusrelation}
\end{figure}

In the end, we substitute the three hybrid EOSs into the TOV equations and integrate them to get the M-R relation of the hybrid stars which is shown in Fig.~\ref{Fig:massradiusrelation}. In this Figure, when $\langle\bar{u}u\rangle|_{T=0, \mu=0} =-(230$ MeV$)^3$, $-(250$ MeV$)^3$, $-(270$ MeV$)^3$, the maximum mass of the hybrid stars is 1.968, 1.893, 1.851 times of the solar mass with the radii of 10.85 km, 10.93 km and 11.15 km respectively (According to our calculation, the quark matter core are about 1.32,
1.46, 1.51 times the solar mass, with the radius of 8.2 km, 8.97 km and 9.26 km respectively.). It's noted that the result confirms the recent astro observation PSR J0348+0432, PSR J1614-2230, PSR J1946+3417 that the pulsars are measured the $2.01 \pm 0.04, 1.928 \pm 0.017 ,1.828 \pm 0.22$ solar mass respectively ~\cite{Antoniadis1233232,Fonseca:2016tux,doi:10.1093/mnras/stw2947,doi:10.1146/annurev-astro-081915-023322}. And we can see that the maximum mass in these three cases do not differ too much, which illustrate that our model and result are parameter-insensitive. By the way, there is not only one interpolation method to get the EOS of hybrid stars~\cite{PhysRevD.91.045003}, however, a different interpolation method doesn't make a big difference because the interpolation function should be smooth at the boundaries of the interpolating interval. So in the crossover region, the EOSs of the hybrid stars with different interpolation methods will not differ widely.
\section{Summary and discussion}\label{three}
In this paper, we introduce the EOS of 2+1 flavors NJL model with PTR, then choose three sets of parameters to calculate the EOS respectively for comparison, namely, the parameters fixed with $u$ quark condensate of $-(230$ MeV$)^3$, $-(250$ MeV$)^3$ and $-(270$ MeV$)^3$ at zero chemical potential and zero temperature. After a series of calculations, we obtain the EOS with a property of crossover in the phase transition region. Then combined with APR model of the hadronic matter and an interpolation method in the crossover region, we get three hybrid EOSs corresponding to three different parameter sets. Finally, the TOV equations are integrated and the M-R relation is obtained. By analysis, the maximum masses corresponding to three hybrid EOSs don't have much difference, which reflects that the choices of the parameters don't play an important role and the result is solid. On the other hand, although the EOS we introduce in this paper is relatively soft, the maximum mass of hybrid stars we deduce is around 1.9 times of solar mass. Above all, it's significant that our study accords with the new measurement and progress in astro study that the neutron-star mass distribution is much wider than previously thought, with three known pulsars now firmly in the 1.9-2.0 solar mass range~\cite{Antoniadis1233232,Fonseca:2016tux,doi:10.1093/mnras/stw2947} and the radii in the 10-11.5 km range~\cite{doi:10.1146/annurev-astro-081915-023322}.

\acknowledgments
This work is supported by the National Natural Science Foundation of China (under Grants No. 11475085, No. 11535005, and No. 11690030).

\bibliography{reference}

\begin{thebibliography}{44}%
\makeatletter
\providecommand \@ifxundefined [1]{%
 \@ifx{#1\undefined}
}%
\providecommand \@ifnum [1]{%
 \ifnum #1\expandafter \@firstoftwo
 \else \expandafter \@secondoftwo
 \fi
}%
\providecommand \@ifx [1]{%
 \ifx #1\expandafter \@firstoftwo
 \else \expandafter \@secondoftwo
 \fi
}%
\providecommand \natexlab [1]{#1}%
\providecommand \enquote  [1]{``#1''}%
\providecommand \bibnamefont  [1]{#1}%
\providecommand \bibfnamefont [1]{#1}%
\providecommand \citenamefont [1]{#1}%
\providecommand \href@noop [0]{\@secondoftwo}%
\providecommand \href [0]{\begingroup \@sanitize@url \@href}%
\providecommand \@href[1]{\@@startlink{#1}\@@href}%
\providecommand \@@href[1]{\endgroup#1\@@endlink}%
\providecommand \@sanitize@url [0]{\catcode `\\12\catcode `\$12\catcode
  `\&12\catcode `\#12\catcode `\^12\catcode `\_12\catcode `\%12\relax}%
\providecommand \@@startlink[1]{}%
\providecommand \@@endlink[0]{}%
\providecommand \url  [0]{\begingroup\@sanitize@url \@url }%
\providecommand \@url [1]{\endgroup\@href {#1}{\urlprefix }}%
\providecommand \urlprefix  [0]{URL }%
\providecommand \Eprint [0]{\href }%
\providecommand \doibase [0]{http://dx.doi.org/}%
\providecommand \selectlanguage [0]{\@gobble}%
\providecommand \bibinfo  [0]{\@secondoftwo}%
\providecommand \bibfield  [0]{\@secondoftwo}%
\providecommand \translation [1]{[#1]}%
\providecommand \BibitemOpen [0]{}%
\providecommand \bibitemStop [0]{}%
\providecommand \bibitemNoStop [0]{.\EOS\space}%
\providecommand \EOS [0]{\spacefactor3000\relax}%
\providecommand \BibitemShut  [1]{\csname bibitem#1\endcsname}%
\let\auto@bib@innerbib\@empty
\bibitem [{\citenamefont {Borsanyi}\ \emph {et~al.}(2010)\citenamefont
  {Borsanyi}, \citenamefont {Fodor}, \citenamefont {Hoelbling}, \citenamefont
  {Katz}, \citenamefont {Krieg}, \citenamefont {Ratti},\ and\ \citenamefont
  {Szabo}}]{Borsanyi:2010bpa}%
  \BibitemOpen
  \bibfield  {author} {\bibinfo {author} {\bibfnamefont {S.}~\bibnamefont
  {Borsanyi}}, \bibinfo {author} {\bibfnamefont {Z.}~\bibnamefont {Fodor}},
  \bibinfo {author} {\bibfnamefont {C.}~\bibnamefont {Hoelbling}}, \bibinfo
  {author} {\bibfnamefont {S.~D.}\ \bibnamefont {Katz}}, \bibinfo {author}
  {\bibfnamefont {S.}~\bibnamefont {Krieg}}, \bibinfo {author} {\bibfnamefont
  {C.}~\bibnamefont {Ratti}}, \ and\ \bibinfo {author} {\bibfnamefont {K.~K.}\
  \bibnamefont {Szabo}} (\bibinfo {collaboration} {Wuppertal-Budapest}),\
  }\href {\doibase 10.1007/JHEP09(2010)073} {\bibfield  {journal} {\bibinfo
  {journal} {JHEP}\ }\textbf {\bibinfo {volume} {09}},\ \bibinfo {pages} {073}
  (\bibinfo {year} {2010})},\ \Eprint {http://arxiv.org/abs/1005.3508}
  {arXiv:1005.3508 [hep-lat]} \BibitemShut {NoStop}%
\bibitem [{\citenamefont {Ejiri}\ and\ \citenamefont
  {Yamada}(2013)}]{PhysRevLett.110.172001}%
  \BibitemOpen
  \bibfield  {author} {\bibinfo {author} {\bibfnamefont {S.}~\bibnamefont
  {Ejiri}}\ and\ \bibinfo {author} {\bibfnamefont {N.}~\bibnamefont {Yamada}},\
  }\href {\doibase 10.1103/PhysRevLett.110.172001} {\bibfield  {journal}
  {\bibinfo  {journal} {Phys. Rev. Lett.}\ }\textbf {\bibinfo {volume} {110}},\
  \bibinfo {pages} {172001} (\bibinfo {year} {2013})}\BibitemShut {NoStop}%
\bibitem [{\citenamefont {Klevansky}(1992)}]{RevModPhys.64.649}%
  \BibitemOpen
  \bibfield  {author} {\bibinfo {author} {\bibfnamefont {S.~P.}\ \bibnamefont
  {Klevansky}},\ }\href {\doibase 10.1103/RevModPhys.64.649} {\bibfield
  {journal} {\bibinfo  {journal} {Rev. Mod. Phys.}\ }\textbf {\bibinfo {volume}
  {64}},\ \bibinfo {pages} {649} (\bibinfo {year} {1992})}\BibitemShut
  {NoStop}%
\bibitem [{\citenamefont {Buballa}(2005)}]{Buballa2005205}%
  \BibitemOpen
  \bibfield  {author} {\bibinfo {author} {\bibfnamefont {M.}~\bibnamefont
  {Buballa}},\ }\href {\doibase
  http://dx.doi.org/10.1016/j.physrep.2004.11.004} {\bibfield  {journal}
  {\bibinfo  {journal} {Phys. Rep.}\ }\textbf {\bibinfo {volume} {407}},\
  \bibinfo {pages} {205 } (\bibinfo {year} {2005})}\BibitemShut {NoStop}%
\bibitem [{\citenamefont {Cui}\ \emph {et~al.}(2013)\citenamefont {Cui},
  \citenamefont {Shi}, \citenamefont {Xia}, \citenamefont {Jiang},\ and\
  \citenamefont {Zong}}]{Cui:2013tva}%
  \BibitemOpen
  \bibfield  {author} {\bibinfo {author} {\bibfnamefont {Z.-F.}\ \bibnamefont
  {Cui}}, \bibinfo {author} {\bibfnamefont {C.}~\bibnamefont {Shi}}, \bibinfo
  {author} {\bibfnamefont {Y.-H.}\ \bibnamefont {Xia}}, \bibinfo {author}
  {\bibfnamefont {Y.}~\bibnamefont {Jiang}}, \ and\ \bibinfo {author}
  {\bibfnamefont {H.-S.}\ \bibnamefont {Zong}},\ }\href {\doibase
  10.1140/epjc/s10052-013-2612-6} {\bibfield  {journal} {\bibinfo  {journal}
  {Eur. Phys. J.}\ }\textbf {\bibinfo {volume} {C73}},\ \bibinfo {pages} {2612}
  (\bibinfo {year} {2013})}\BibitemShut {NoStop}%
\bibitem [{\citenamefont {Cui}\ \emph {et~al.}(2014{\natexlab{a}})\citenamefont
  {Cui}, \citenamefont {Shi}, \citenamefont {Sun}, \citenamefont {Wang},\ and\
  \citenamefont {Zong}}]{Cui:2013aba}%
  \BibitemOpen
  \bibfield  {author} {\bibinfo {author} {\bibfnamefont {Z.-f.}\ \bibnamefont
  {Cui}}, \bibinfo {author} {\bibfnamefont {C.}~\bibnamefont {Shi}}, \bibinfo
  {author} {\bibfnamefont {W.-m.}\ \bibnamefont {Sun}}, \bibinfo {author}
  {\bibfnamefont {Y.-l.}\ \bibnamefont {Wang}}, \ and\ \bibinfo {author}
  {\bibfnamefont {H.-s.}\ \bibnamefont {Zong}},\ }\href {\doibase
  10.1140/epjc/s10052-014-2782-x} {\bibfield  {journal} {\bibinfo  {journal}
  {Eur. Phys. J.}\ }\textbf {\bibinfo {volume} {C74}},\ \bibinfo {pages} {2782}
  (\bibinfo {year} {2014}{\natexlab{a}})},\ \Eprint
  {http://arxiv.org/abs/1311.4014} {arXiv:1311.4014 [hep-ph]} \BibitemShut
  {NoStop}%
\bibitem [{\citenamefont {Kohyama}\ \emph {et~al.}(2015)\citenamefont
  {Kohyama}, \citenamefont {Kimura},\ and\ \citenamefont
  {Inagaki}}]{NuclPhysB.896.682}%
  \BibitemOpen
  \bibfield  {author} {\bibinfo {author} {\bibfnamefont {H.}~\bibnamefont
  {Kohyama}}, \bibinfo {author} {\bibfnamefont {D.}~\bibnamefont {Kimura}}, \
  and\ \bibinfo {author} {\bibfnamefont {T.}~\bibnamefont {Inagaki}},\
  }\href@noop {} {\bibfield  {journal} {\bibinfo  {journal} {Nucl. Phys. B}\
  }\textbf {\bibinfo {volume} {896}},\ \bibinfo {pages} {682} (\bibinfo {year}
  {2015})}\BibitemShut {NoStop}%
\bibitem [{\citenamefont {Roberts}\ and\ \citenamefont
  {Williams}(1994)}]{ROBERTS1994477}%
  \BibitemOpen
  \bibfield  {author} {\bibinfo {author} {\bibfnamefont {C.~D.}\ \bibnamefont
  {Roberts}}\ and\ \bibinfo {author} {\bibfnamefont {A.~G.}\ \bibnamefont
  {Williams}},\ }\href {\doibase
  http://dx.doi.org/10.1016/0146-6410(94)90049-3} {\bibfield  {journal}
  {\bibinfo  {journal} {Prog. Part. Nucl. Phys.}\ }\textbf {\bibinfo {volume}
  {33}},\ \bibinfo {pages} {477 } (\bibinfo {year} {1994})}\BibitemShut
  {NoStop}%
\bibitem [{\citenamefont {Roberts}\ and\ \citenamefont
  {Schmidt}(2000)}]{Roberts2000S1}%
  \BibitemOpen
  \bibfield  {author} {\bibinfo {author} {\bibfnamefont {C.}~\bibnamefont
  {Roberts}}\ and\ \bibinfo {author} {\bibfnamefont {S.}~\bibnamefont
  {Schmidt}},\ }\href {\doibase
  http://dx.doi.org/10.1016/S0146-6410(00)90011-5} {\bibfield  {journal}
  {\bibinfo  {journal} {Prog. Part. Nucl. Phys.}\ }\textbf {\bibinfo {volume}
  {45, Supplement 1}},\ \bibinfo {pages} {S1 } (\bibinfo {year}
  {2000})}\BibitemShut {NoStop}%
\bibitem [{\citenamefont {Maris}\ and\ \citenamefont
  {Roberts}(2003)}]{doi:10.1142/S0218301303001326}%
  \BibitemOpen
  \bibfield  {author} {\bibinfo {author} {\bibfnamefont {P.}~\bibnamefont
  {Maris}}\ and\ \bibinfo {author} {\bibfnamefont {C.~D.}\ \bibnamefont
  {Roberts}},\ }\href {\doibase 10.1142/S0218301303001326} {\bibfield
  {journal} {\bibinfo  {journal} {Int. J. Mod. Phys. E}\ }\textbf {\bibinfo
  {volume} {12}},\ \bibinfo {pages} {297} (\bibinfo {year} {2003})}\BibitemShut
  {NoStop}%
\bibitem [{\citenamefont {Cl{\"o}et}\ and\ \citenamefont
  {Roberts}(2014)}]{Cloet20141}%
  \BibitemOpen
  \bibfield  {author} {\bibinfo {author} {\bibfnamefont {I.~C.}\ \bibnamefont
  {Cl{\"o}et}}\ and\ \bibinfo {author} {\bibfnamefont {C.~D.}\ \bibnamefont
  {Roberts}},\ }\href {\doibase http://dx.doi.org/10.1016/j.ppnp.2014.02.001}
  {\bibfield  {journal} {\bibinfo  {journal} {Prog. Part. Nucl. Phys.}\
  }\textbf {\bibinfo {volume} {77}},\ \bibinfo {pages} {1 } (\bibinfo {year}
  {2014})}\BibitemShut {NoStop}%
\bibitem [{\citenamefont {Zhao}\ \emph {et~al.}(2014)\citenamefont {Zhao},
  \citenamefont {Cui}, \citenamefont {Jiang},\ and\ \citenamefont
  {Zong}}]{PhysRevD.90.114031}%
  \BibitemOpen
  \bibfield  {author} {\bibinfo {author} {\bibfnamefont {A.-M.}\ \bibnamefont
  {Zhao}}, \bibinfo {author} {\bibfnamefont {Z.-F.}\ \bibnamefont {Cui}},
  \bibinfo {author} {\bibfnamefont {Y.}~\bibnamefont {Jiang}}, \ and\ \bibinfo
  {author} {\bibfnamefont {H.-S.}\ \bibnamefont {Zong}},\ }\href {\doibase
  10.1103/PhysRevD.90.114031} {\bibfield  {journal} {\bibinfo  {journal} {Phys.
  Rev. D}\ }\textbf {\bibinfo {volume} {90}},\ \bibinfo {pages} {114031}
  (\bibinfo {year} {2014})}\BibitemShut {NoStop}%
\bibitem [{\citenamefont {Wang}\ \emph {et~al.}(2015)\citenamefont {Wang},
  \citenamefont {Wang}, \citenamefont {Cui},\ and\ \citenamefont
  {Zong}}]{PhysRevD.91.034017}%
  \BibitemOpen
  \bibfield  {author} {\bibinfo {author} {\bibfnamefont {B.}~\bibnamefont
  {Wang}}, \bibinfo {author} {\bibfnamefont {Y.-L.}\ \bibnamefont {Wang}},
  \bibinfo {author} {\bibfnamefont {Z.-F.}\ \bibnamefont {Cui}}, \ and\
  \bibinfo {author} {\bibfnamefont {H.-S.}\ \bibnamefont {Zong}},\ }\href
  {\doibase 10.1103/PhysRevD.91.034017} {\bibfield  {journal} {\bibinfo
  {journal} {Phys. Rev. D}\ }\textbf {\bibinfo {volume} {91}},\ \bibinfo
  {pages} {034017} (\bibinfo {year} {2015})}\BibitemShut {NoStop}%
\bibitem [{\citenamefont {Xu}\ \emph {et~al.}(2015)\citenamefont {Xu},
  \citenamefont {Cui}, \citenamefont {Wang}, \citenamefont {Shi}, \citenamefont
  {Yang},\ and\ \citenamefont {Zong}}]{PhysRevD.91.056003}%
  \BibitemOpen
  \bibfield  {author} {\bibinfo {author} {\bibfnamefont {S.-S.}\ \bibnamefont
  {Xu}}, \bibinfo {author} {\bibfnamefont {Z.-F.}\ \bibnamefont {Cui}},
  \bibinfo {author} {\bibfnamefont {B.}~\bibnamefont {Wang}}, \bibinfo {author}
  {\bibfnamefont {Y.-M.}\ \bibnamefont {Shi}}, \bibinfo {author} {\bibfnamefont
  {Y.-C.}\ \bibnamefont {Yang}}, \ and\ \bibinfo {author} {\bibfnamefont
  {H.-S.}\ \bibnamefont {Zong}},\ }\href {\doibase 10.1103/PhysRevD.91.056003}
  {\bibfield  {journal} {\bibinfo  {journal} {Phys. Rev. D}\ }\textbf {\bibinfo
  {volume} {91}},\ \bibinfo {pages} {056003} (\bibinfo {year}
  {2015})}\BibitemShut {NoStop}%
\bibitem [{\citenamefont {Pisarski}(1984)}]{PhysRevD.29.2423}%
  \BibitemOpen
  \bibfield  {author} {\bibinfo {author} {\bibfnamefont {R.~D.}\ \bibnamefont
  {Pisarski}},\ }\href {\doibase 10.1103/PhysRevD.29.2423} {\bibfield
  {journal} {\bibinfo  {journal} {Phys. Rev. D}\ }\textbf {\bibinfo {volume}
  {29}},\ \bibinfo {pages} {2423} (\bibinfo {year} {1984})}\BibitemShut
  {NoStop}%
\bibitem [{\citenamefont {Yin}\ \emph {et~al.}(2014)\citenamefont {Yin},
  \citenamefont {Shi}, \citenamefont {Cui}, \citenamefont {Feng},\ and\
  \citenamefont {Zong}}]{PhysRevD.90.036007}%
  \BibitemOpen
  \bibfield  {author} {\bibinfo {author} {\bibfnamefont {P.-l.}\ \bibnamefont
  {Yin}}, \bibinfo {author} {\bibfnamefont {Y.-m.}\ \bibnamefont {Shi}},
  \bibinfo {author} {\bibfnamefont {Z.-f.}\ \bibnamefont {Cui}}, \bibinfo
  {author} {\bibfnamefont {H.-t.}\ \bibnamefont {Feng}}, \ and\ \bibinfo
  {author} {\bibfnamefont {H.-s.}\ \bibnamefont {Zong}},\ }\href {\doibase
  10.1103/PhysRevD.90.036007} {\bibfield  {journal} {\bibinfo  {journal} {Phys.
  Rev. D}\ }\textbf {\bibinfo {volume} {90}},\ \bibinfo {pages} {036007}
  (\bibinfo {year} {2014})}\BibitemShut {NoStop}%
\bibitem [{\citenamefont {Li}\ \emph {et~al.}(2014)\citenamefont {Li},
  \citenamefont {Hou}, \citenamefont {Cui}, \citenamefont {Feng}, \citenamefont
  {Jiang},\ and\ \citenamefont {Zong}}]{PhysRevD.90.073013}%
  \BibitemOpen
  \bibfield  {author} {\bibinfo {author} {\bibfnamefont {J.-F.}\ \bibnamefont
  {Li}}, \bibinfo {author} {\bibfnamefont {F.-Y.}\ \bibnamefont {Hou}},
  \bibinfo {author} {\bibfnamefont {Z.-F.}\ \bibnamefont {Cui}}, \bibinfo
  {author} {\bibfnamefont {H.-T.}\ \bibnamefont {Feng}}, \bibinfo {author}
  {\bibfnamefont {Y.}~\bibnamefont {Jiang}}, \ and\ \bibinfo {author}
  {\bibfnamefont {H.-S.}\ \bibnamefont {Zong}},\ }\href {\doibase
  10.1103/PhysRevD.90.073013} {\bibfield  {journal} {\bibinfo  {journal} {Phys.
  Rev. D}\ }\textbf {\bibinfo {volume} {90}},\ \bibinfo {pages} {073013}
  (\bibinfo {year} {2014})}\BibitemShut {NoStop}%
\bibitem [{\citenamefont {Nielsen}(1985)}]{NIELSEN1985401}%
  \BibitemOpen
  \bibfield  {author} {\bibinfo {author} {\bibfnamefont {N.}~\bibnamefont
  {Nielsen}},\ }\href {\doibase http://dx.doi.org/10.1016/0550-3213(85)90455-9}
  {\bibfield  {journal} {\bibinfo  {journal} {Nucl. Phys. B}\ }\textbf
  {\bibinfo {volume} {252}},\ \bibinfo {pages} {401 } (\bibinfo {year}
  {1985})}\BibitemShut {NoStop}%
\bibitem [{\citenamefont {Cui}\ \emph {et~al.}(2014{\natexlab{b}})\citenamefont
  {Cui}, \citenamefont {Du},\ and\ \citenamefont {Zong}}]{Cui:2014hya}%
  \BibitemOpen
  \bibfield  {author} {\bibinfo {author} {\bibfnamefont {Z.-F.}\ \bibnamefont
  {Cui}}, \bibinfo {author} {\bibfnamefont {Y.-L.}\ \bibnamefont {Du}}, \ and\
  \bibinfo {author} {\bibfnamefont {H.-S.}\ \bibnamefont {Zong}},\ }\bibfield
  {booktitle} {\emph {\bibinfo {booktitle} {{Proceedings, Workshop on Hadron
  Nuclear Physics (HNP 2013): Zhangjiajie, China, July 18-22, 2013}}},\ }\href
  {\doibase 10.1142/S2010194514602324} {\bibfield  {journal} {\bibinfo
  {journal} {Int. J. Mod. Phys. Conf. Ser.}\ }\textbf {\bibinfo {volume}
  {29}},\ \bibinfo {pages} {1460232} (\bibinfo {year}
  {2014}{\natexlab{b}})}\BibitemShut {NoStop}%
\bibitem [{\citenamefont {Zhang}\ \emph {et~al.}(2016)\citenamefont {Zhang},
  \citenamefont {Shi}, \citenamefont {Xu},\ and\ \citenamefont
  {Zong}}]{doi:10.1142/S0217732316500863}%
  \BibitemOpen
  \bibfield  {author} {\bibinfo {author} {\bibfnamefont {J.-L.}\ \bibnamefont
  {Zhang}}, \bibinfo {author} {\bibfnamefont {Y.-M.}\ \bibnamefont {Shi}},
  \bibinfo {author} {\bibfnamefont {S.-S.}\ \bibnamefont {Xu}}, \ and\ \bibinfo
  {author} {\bibfnamefont {H.-S.}\ \bibnamefont {Zong}},\ }\href {\doibase
  10.1142/S0217732316500863} {\bibfield  {journal} {\bibinfo  {journal} {Mod.
  Phys. Lett. A}\ }\textbf {\bibinfo {volume} {31}},\ \bibinfo {pages}
  {1650086} (\bibinfo {year} {2016})}\BibitemShut {NoStop}%
\bibitem [{\citenamefont {Li}\ \emph {et~al.}(2015)\citenamefont {Li},
  \citenamefont {Zuo},\ and\ \citenamefont {Peng}}]{PhysRevC.91.035803}%
  \BibitemOpen
  \bibfield  {author} {\bibinfo {author} {\bibfnamefont {A.}~\bibnamefont
  {Li}}, \bibinfo {author} {\bibfnamefont {W.}~\bibnamefont {Zuo}}, \ and\
  \bibinfo {author} {\bibfnamefont {G.~X.}\ \bibnamefont {Peng}},\ }\href
  {\doibase 10.1103/PhysRevC.91.035803} {\bibfield  {journal} {\bibinfo
  {journal} {Phys. Rev. C}\ }\textbf {\bibinfo {volume} {91}},\ \bibinfo
  {pages} {035803} (\bibinfo {year} {2015})}\BibitemShut {NoStop}%
\bibitem [{\citenamefont {Mallick}\ and\ \citenamefont
  {Sahu}(2014)}]{Mallick201496}%
  \BibitemOpen
  \bibfield  {author} {\bibinfo {author} {\bibfnamefont {R.}~\bibnamefont
  {Mallick}}\ and\ \bibinfo {author} {\bibfnamefont {P.}~\bibnamefont {Sahu}},\
  }\href {\doibase http://dx.doi.org/10.1016/j.nuclphysa.2013.11.009}
  {\bibfield  {journal} {\bibinfo  {journal} {Nucl. Phys. A}\ }\textbf
  {\bibinfo {volume} {921}},\ \bibinfo {pages} {96 } (\bibinfo {year}
  {2014})}\BibitemShut {NoStop}%
\bibitem [{\citenamefont {Benic}\ \emph {et~al.}(2015)\citenamefont {Benic},
  \citenamefont {Blaschke}, \citenamefont {Alvarez-Castillo}, \citenamefont
  {Fischer},\ and\ \citenamefont {Typel}}]{refId0}%
  \BibitemOpen
  \bibfield  {author} {\bibinfo {author} {\bibfnamefont {S.}~\bibnamefont
  {Benic}}, \bibinfo {author} {\bibfnamefont {D.}~\bibnamefont {Blaschke}},
  \bibinfo {author} {\bibfnamefont {D.~E.}\ \bibnamefont {Alvarez-Castillo}},
  \bibinfo {author} {\bibfnamefont {T.}~\bibnamefont {Fischer}}, \ and\
  \bibinfo {author} {\bibfnamefont {S.}~\bibnamefont {Typel}},\ }\href
  {\doibase 10.1051/0004-6361/201425318} {\bibfield  {journal} {\bibinfo
  {journal} {A\&A}\ }\textbf {\bibinfo {volume} {577}},\ \bibinfo {pages} {A40}
  (\bibinfo {year} {2015})}\BibitemShut {NoStop}%
\bibitem [{\citenamefont {Alvarez-Castillo}\ and\ \citenamefont
  {Blaschke}(2015)}]{Alvarez-Castillo2015}%
  \BibitemOpen
  \bibfield  {author} {\bibinfo {author} {\bibfnamefont {D.~E.}\ \bibnamefont
  {Alvarez-Castillo}}\ and\ \bibinfo {author} {\bibfnamefont {D.}~\bibnamefont
  {Blaschke}},\ }\href {\doibase 10.1134/S1063779615050032} {\bibfield
  {journal} {\bibinfo  {journal} {Phys. Part. Nucl.}\ }\textbf {\bibinfo
  {volume} {46}},\ \bibinfo {pages} {846} (\bibinfo {year} {2015})}\BibitemShut
  {NoStop}%
\bibitem [{\citenamefont {de~Forcrand}\ \emph {et~al.}(2014)\citenamefont
  {de~Forcrand}, \citenamefont {Langelage}, \citenamefont {Philipsen},\ and\
  \citenamefont {Unger}}]{PhysRevLett.113.152002}%
  \BibitemOpen
  \bibfield  {author} {\bibinfo {author} {\bibfnamefont {P.}~\bibnamefont
  {de~Forcrand}}, \bibinfo {author} {\bibfnamefont {J.}~\bibnamefont
  {Langelage}}, \bibinfo {author} {\bibfnamefont {O.}~\bibnamefont
  {Philipsen}}, \ and\ \bibinfo {author} {\bibfnamefont {W.}~\bibnamefont
  {Unger}},\ }\href {\doibase 10.1103/PhysRevLett.113.152002} {\bibfield
  {journal} {\bibinfo  {journal} {Phys. Rev. Lett.}\ }\textbf {\bibinfo
  {volume} {113}},\ \bibinfo {pages} {152002} (\bibinfo {year}
  {2014})}\BibitemShut {NoStop}%
\bibitem [{\citenamefont {Endrodi}(2015)}]{Endrodi:2015oba}%
  \BibitemOpen
  \bibfield  {author} {\bibinfo {author} {\bibfnamefont {G.}~\bibnamefont
  {Endrodi}},\ }\href {\doibase 10.1007/JHEP07(2015)173} {\bibfield  {journal}
  {\bibinfo  {journal} {JHEP}\ }\textbf {\bibinfo {volume} {07}},\ \bibinfo
  {pages} {173} (\bibinfo {year} {2015})},\ \Eprint
  {http://arxiv.org/abs/1504.08280} {arXiv:1504.08280 [hep-lat]} \BibitemShut
  {NoStop}%
\bibitem [{\citenamefont {Braguta}\ \emph {et~al.}(2015)\citenamefont
  {Braguta}, \citenamefont {Goy}, \citenamefont {Ilgenfritz}, \citenamefont
  {Kotov}, \citenamefont {Molochkov}, \citenamefont {Muller-Preussker},\ and\
  \citenamefont {Petersson}}]{Braguta:2015zta}%
  \BibitemOpen
  \bibfield  {author} {\bibinfo {author} {\bibfnamefont {V.~V.}\ \bibnamefont
  {Braguta}}, \bibinfo {author} {\bibfnamefont {V.~A.}\ \bibnamefont {Goy}},
  \bibinfo {author} {\bibfnamefont {E.~M.}\ \bibnamefont {Ilgenfritz}},
  \bibinfo {author} {\bibfnamefont {A.~{\relax Yu}.}\ \bibnamefont {Kotov}},
  \bibinfo {author} {\bibfnamefont {A.~V.}\ \bibnamefont {Molochkov}}, \bibinfo
  {author} {\bibfnamefont {M.}~\bibnamefont {Muller-Preussker}}, \ and\
  \bibinfo {author} {\bibfnamefont {B.}~\bibnamefont {Petersson}},\ }\href
  {\doibase 10.1007/JHEP06(2015)094} {\bibfield  {journal} {\bibinfo  {journal}
  {JHEP}\ }\textbf {\bibinfo {volume} {06}},\ \bibinfo {pages} {094} (\bibinfo
  {year} {2015})},\ \Eprint {http://arxiv.org/abs/1503.06670} {arXiv:1503.06670
  [hep-lat]} \BibitemShut {NoStop}%
\bibitem [{\citenamefont {Bratovic}\ \emph {et~al.}(2013)\citenamefont
  {Bratovic}, \citenamefont {Hatsuda},\ and\ \citenamefont
  {Weise}}]{Bratovic2013131}%
  \BibitemOpen
  \bibfield  {author} {\bibinfo {author} {\bibfnamefont {N.}~\bibnamefont
  {Bratovic}}, \bibinfo {author} {\bibfnamefont {T.}~\bibnamefont {Hatsuda}}, \
  and\ \bibinfo {author} {\bibfnamefont {W.}~\bibnamefont {Weise}},\ }\href
  {\doibase http://dx.doi.org/10.1016/j.physletb.2013.01.003} {\bibfield
  {journal} {\bibinfo  {journal} {Phys. Lett. B}\ }\textbf {\bibinfo {volume}
  {719}},\ \bibinfo {pages} {131 } (\bibinfo {year} {2013})}\BibitemShut
  {NoStop}%
\bibitem [{\citenamefont {Masuda}\ \emph
  {et~al.}(2013{\natexlab{a}})\citenamefont {Masuda}, \citenamefont {Hatsuda},\
  and\ \citenamefont {Takatsuka}}]{Masuda:2012ed}%
  \BibitemOpen
  \bibfield  {author} {\bibinfo {author} {\bibfnamefont {K.}~\bibnamefont
  {Masuda}}, \bibinfo {author} {\bibfnamefont {T.}~\bibnamefont {Hatsuda}}, \
  and\ \bibinfo {author} {\bibfnamefont {T.}~\bibnamefont {Takatsuka}},\ }\href
  {\doibase 10.1093/ptep/ptt045} {\bibfield  {journal} {\bibinfo  {journal}
  {PTEP}\ }\textbf {\bibinfo {volume} {2013}},\ \bibinfo {pages} {073D01}
  (\bibinfo {year} {2013}{\natexlab{a}})},\ \Eprint
  {http://arxiv.org/abs/1212.6803} {arXiv:1212.6803 [nucl-th]} \BibitemShut
  {NoStop}%
\bibitem [{\citenamefont {Masuda}\ \emph
  {et~al.}(2013{\natexlab{b}})\citenamefont {Masuda}, \citenamefont {Hatsuda},\
  and\ \citenamefont {Takatsuka}}]{0004-637X-764-1-12}%
  \BibitemOpen
  \bibfield  {author} {\bibinfo {author} {\bibfnamefont {K.}~\bibnamefont
  {Masuda}}, \bibinfo {author} {\bibfnamefont {T.}~\bibnamefont {Hatsuda}}, \
  and\ \bibinfo {author} {\bibfnamefont {T.}~\bibnamefont {Takatsuka}},\ }\href
  {http://stacks.iop.org/0004-637X/764/i=1/a=12} {\bibfield  {journal}
  {\bibinfo  {journal} {Astrophys. J}\ }\textbf {\bibinfo {volume} {764}},\
  \bibinfo {pages} {12} (\bibinfo {year} {2013}{\natexlab{b}})}\BibitemShut
  {NoStop}%
\bibitem [{\citenamefont {Kojo}\ \emph {et~al.}(2015)\citenamefont {Kojo},
  \citenamefont {Powell}, \citenamefont {Song},\ and\ \citenamefont
  {Baym}}]{PhysRevD.91.045003}%
  \BibitemOpen
  \bibfield  {author} {\bibinfo {author} {\bibfnamefont {T.}~\bibnamefont
  {Kojo}}, \bibinfo {author} {\bibfnamefont {P.~D.}\ \bibnamefont {Powell}},
  \bibinfo {author} {\bibfnamefont {Y.}~\bibnamefont {Song}}, \ and\ \bibinfo
  {author} {\bibfnamefont {G.}~\bibnamefont {Baym}},\ }\href {\doibase
  10.1103/PhysRevD.91.045003} {\bibfield  {journal} {\bibinfo  {journal} {Phys.
  Rev. D}\ }\textbf {\bibinfo {volume} {91}},\ \bibinfo {pages} {045003}
  (\bibinfo {year} {2015})}\BibitemShut {NoStop}%
\bibitem [{\citenamefont {Zhao}\ \emph {et~al.}()\citenamefont {Zhao},
  \citenamefont {Li},\ and\ \citenamefont {Zong}}]{ZhaoYP}%
  \BibitemOpen
  \bibfield  {author} {\bibinfo {author} {\bibfnamefont {Y.-P.}\ \bibnamefont
  {Zhao}}, \bibinfo {author} {\bibfnamefont {C.-M.}\ \bibnamefont {Li}}, \ and\
  \bibinfo {author} {\bibfnamefont {H.-S.}\ \bibnamefont {Zong}},\ }\href@noop
  {} {\bibinfo  {journal} {(unpublished)}\ }\BibitemShut {NoStop}%
\bibitem [{\citenamefont {Antoniadis}\ \emph {et~al.}(2013)\citenamefont
  {Antoniadis}, \citenamefont {Freire}, \citenamefont {Wex}, \citenamefont
  {Tauris}, \citenamefont {Lynch}, \citenamefont {van Kerkwijk}, \citenamefont
  {Kramer}, \citenamefont {Bassa}, \citenamefont {Dhillon}, \citenamefont
  {Driebe}, \citenamefont {Hessels}, \citenamefont {Kaspi}, \citenamefont
  {Kondratiev}, \citenamefont {Langer}, \citenamefont {Marsh}, \citenamefont
  {McLaughlin}, \citenamefont {Pennucci}, \citenamefont {Ransom}, \citenamefont
  {Stairs}, \citenamefont {van Leeuwen}, \citenamefont {Verbiest},\ and\
  \citenamefont {Whelan}}]{Antoniadis1233232}%
  \BibitemOpen
\bibfield  {journal} {  }\bibfield  {author} {\bibinfo {author} {\bibfnamefont
  {J.}~\bibnamefont {Antoniadis}}, \bibinfo {author} {\bibfnamefont {P.~C.~C.}\
  \bibnamefont {Freire}}, \bibinfo {author} {\bibfnamefont {N.}~\bibnamefont
  {Wex}}, \bibinfo {author} {\bibfnamefont {T.~M.}\ \bibnamefont {Tauris}},
  \bibinfo {author} {\bibfnamefont {R.~S.}\ \bibnamefont {Lynch}}, \bibinfo
  {author} {\bibfnamefont {M.~H.}\ \bibnamefont {van Kerkwijk}}, \bibinfo
  {author} {\bibfnamefont {M.}~\bibnamefont {Kramer}}, \bibinfo {author}
  {\bibfnamefont {C.}~\bibnamefont {Bassa}}, \bibinfo {author} {\bibfnamefont
  {V.~S.}\ \bibnamefont {Dhillon}}, \bibinfo {author} {\bibfnamefont
  {T.}~\bibnamefont {Driebe}}, \bibinfo {author} {\bibfnamefont {J.~W.~T.}\
  \bibnamefont {Hessels}}, \bibinfo {author} {\bibfnamefont {V.~M.}\
  \bibnamefont {Kaspi}}, \bibinfo {author} {\bibfnamefont {V.~I.}\ \bibnamefont
  {Kondratiev}}, \bibinfo {author} {\bibfnamefont {N.}~\bibnamefont {Langer}},
  \bibinfo {author} {\bibfnamefont {T.~R.}\ \bibnamefont {Marsh}}, \bibinfo
  {author} {\bibfnamefont {M.~A.}\ \bibnamefont {McLaughlin}}, \bibinfo
  {author} {\bibfnamefont {T.~T.}\ \bibnamefont {Pennucci}}, \bibinfo {author}
  {\bibfnamefont {S.~M.}\ \bibnamefont {Ransom}}, \bibinfo {author}
  {\bibfnamefont {I.~H.}\ \bibnamefont {Stairs}}, \bibinfo {author}
  {\bibfnamefont {J.}~\bibnamefont {van Leeuwen}}, \bibinfo {author}
  {\bibfnamefont {J.~P.~W.}\ \bibnamefont {Verbiest}}, \ and\ \bibinfo {author}
  {\bibfnamefont {D.~G.}\ \bibnamefont {Whelan}},\ }\href@noop {} {\bibfield
  {journal} {\bibinfo  {journal} {Science}\ }\textbf {\bibinfo {volume} {340}}
  (\bibinfo {year} {2013})}\BibitemShut {NoStop}%
\bibitem [{\citenamefont {Fonseca}\ \emph {et~al.}(2016)\citenamefont {Fonseca}
  \emph {et~al.}}]{Fonseca:2016tux}%
  \BibitemOpen
  \bibfield  {author} {\bibinfo {author} {\bibfnamefont {E.}~\bibnamefont
  {Fonseca}} \emph {et~al.},\ }\href {\doibase 10.3847/0004-637X/832/2/167}
  {\bibfield  {journal} {\bibinfo  {journal} {Astrophys. J.}\ }\textbf
  {\bibinfo {volume} {832}},\ \bibinfo {pages} {167} (\bibinfo {year}
  {2016})},\ \Eprint {http://arxiv.org/abs/1603.00545} {arXiv:1603.00545
  [astro-ph.HE]} \BibitemShut {NoStop}%
\bibitem [{\citenamefont {Barr}\ \emph {et~al.}(2016)\citenamefont {Barr},
  \citenamefont {Freire}, \citenamefont {Kramer}, \citenamefont {Champion},
  \citenamefont {Berezina}, \citenamefont {Bassa}, \citenamefont {Lyne},\ and\
  \citenamefont {Stappers}}]{doi:10.1093/mnras/stw2947}%
  \BibitemOpen
  \bibfield  {author} {\bibinfo {author} {\bibfnamefont {E.~D.}\ \bibnamefont
  {Barr}}, \bibinfo {author} {\bibfnamefont {P.~C.~C.}\ \bibnamefont {Freire}},
  \bibinfo {author} {\bibfnamefont {M.}~\bibnamefont {Kramer}}, \bibinfo
  {author} {\bibfnamefont {D.~J.}\ \bibnamefont {Champion}}, \bibinfo {author}
  {\bibfnamefont {M.}~\bibnamefont {Berezina}}, \bibinfo {author}
  {\bibfnamefont {C.~G.}\ \bibnamefont {Bassa}}, \bibinfo {author}
  {\bibfnamefont {A.~G.}\ \bibnamefont {Lyne}}, \ and\ \bibinfo {author}
  {\bibfnamefont {B.~W.}\ \bibnamefont {Stappers}},\ }\href {\doibase
  10.1093/mnras/stw2947} {\bibfield  {journal} {\bibinfo  {journal} {Monthly
  Notices of the Royal Astronomical Society}\ }\textbf {\bibinfo {volume}
  {465}},\ \bibinfo {pages} {1711} (\bibinfo {year} {2016})}\BibitemShut
  {NoStop}%
\bibitem [{\citenamefont {\"Ozel}\ and\ \citenamefont
  {Freire}(2016)}]{doi:10.1146/annurev-astro-081915-023322}%
  \BibitemOpen
  \bibfield  {author} {\bibinfo {author} {\bibfnamefont {F.}~\bibnamefont
  {\"Ozel}}\ and\ \bibinfo {author} {\bibfnamefont {P.}~\bibnamefont
  {Freire}},\ }\href {\doibase 10.1146/annurev-astro-081915-023322} {\bibfield
  {journal} {\bibinfo  {journal} {Ann. Rev. Astr. Astrophys.}\ }\textbf
  {\bibinfo {volume} {54}},\ \bibinfo {pages} {401} (\bibinfo {year}
  {2016})}\BibitemShut {NoStop}%
\bibitem [{\citenamefont {Zong}\ \emph {et~al.}(2005)\citenamefont {Zong},
  \citenamefont {Chang}, \citenamefont {Hou}, \citenamefont {Sun},\ and\
  \citenamefont {Liu}}]{PhysRevC.71.015205}%
  \BibitemOpen
  \bibfield  {author} {\bibinfo {author} {\bibfnamefont {H.-s.}\ \bibnamefont
  {Zong}}, \bibinfo {author} {\bibfnamefont {L.}~\bibnamefont {Chang}},
  \bibinfo {author} {\bibfnamefont {F.-y.}\ \bibnamefont {Hou}}, \bibinfo
  {author} {\bibfnamefont {W.-m.}\ \bibnamefont {Sun}}, \ and\ \bibinfo
  {author} {\bibfnamefont {Y.-x.}\ \bibnamefont {Liu}},\ }\href {\doibase
  10.1103/PhysRevC.71.015205} {\bibfield  {journal} {\bibinfo  {journal} {Phys.
  Rev. C}\ }\textbf {\bibinfo {volume} {71}},\ \bibinfo {pages} {015205}
  (\bibinfo {year} {2005})}\BibitemShut {NoStop}%
\bibitem [{\citenamefont {Halasz}\ \emph {et~al.}(1998)\citenamefont {Halasz},
  \citenamefont {Jackson}, \citenamefont {Shrock}, \citenamefont {Stephanov},\
  and\ \citenamefont {Verbaarschot}}]{PhysRevD.58.096007}%
  \BibitemOpen
  \bibfield  {author} {\bibinfo {author} {\bibfnamefont {M.~A.}\ \bibnamefont
  {Halasz}}, \bibinfo {author} {\bibfnamefont {A.~D.}\ \bibnamefont {Jackson}},
  \bibinfo {author} {\bibfnamefont {R.~E.}\ \bibnamefont {Shrock}}, \bibinfo
  {author} {\bibfnamefont {M.~A.}\ \bibnamefont {Stephanov}}, \ and\ \bibinfo
  {author} {\bibfnamefont {J.~J.~M.}\ \bibnamefont {Verbaarschot}},\ }\href
  {\doibase 10.1103/PhysRevD.58.096007} {\bibfield  {journal} {\bibinfo
  {journal} {Phys. Rev. D}\ }\textbf {\bibinfo {volume} {58}},\ \bibinfo
  {pages} {096007} (\bibinfo {year} {1998})}\BibitemShut {NoStop}%
\bibitem [{\citenamefont {ZONG}\ and\ \citenamefont
  {SUN}(2008)}]{doi:10.1142/S0217751X08040457}%
  \BibitemOpen
  \bibfield  {author} {\bibinfo {author} {\bibfnamefont {H.-S.}\ \bibnamefont
  {ZONG}}\ and\ \bibinfo {author} {\bibfnamefont {W.-M.}\ \bibnamefont {SUN}},\
  }\href {\doibase 10.1142/S0217751X08040457} {\bibfield  {journal} {\bibinfo
  {journal} {Int. J. Mod. Phys. A}\ }\textbf {\bibinfo {volume} {23}},\
  \bibinfo {pages} {3591} (\bibinfo {year} {2008})}\BibitemShut {NoStop}%
\bibitem [{\citenamefont {Yan}\ \emph {et~al.}(2012)\citenamefont {Yan},
  \citenamefont {Cao}, \citenamefont {Luo}, \citenamefont {Sun},\ and\
  \citenamefont {Zong}}]{PhysRevD.86.114028}%
  \BibitemOpen
  \bibfield  {author} {\bibinfo {author} {\bibfnamefont {Y.}~\bibnamefont
  {Yan}}, \bibinfo {author} {\bibfnamefont {J.}~\bibnamefont {Cao}}, \bibinfo
  {author} {\bibfnamefont {X.-L.}\ \bibnamefont {Luo}}, \bibinfo {author}
  {\bibfnamefont {W.-M.}\ \bibnamefont {Sun}}, \ and\ \bibinfo {author}
  {\bibfnamefont {H.}~\bibnamefont {Zong}},\ }\href {\doibase
  10.1103/PhysRevD.86.114028} {\bibfield  {journal} {\bibinfo  {journal} {Phys.
  Rev. D}\ }\textbf {\bibinfo {volume} {86}},\ \bibinfo {pages} {114028}
  (\bibinfo {year} {2012})}\BibitemShut {NoStop}%
\bibitem [{\citenamefont {Benvenuto}\ and\ \citenamefont
  {Lugones}(1995)}]{PhysRevD.51.1989}%
  \BibitemOpen
  \bibfield  {author} {\bibinfo {author} {\bibfnamefont {O.~G.}\ \bibnamefont
  {Benvenuto}}\ and\ \bibinfo {author} {\bibfnamefont {G.}~\bibnamefont
  {Lugones}},\ }\href {\doibase 10.1103/PhysRevD.51.1989} {\bibfield  {journal}
  {\bibinfo  {journal} {Phys. Rev. D}\ }\textbf {\bibinfo {volume} {51}},\
  \bibinfo {pages} {1989} (\bibinfo {year} {1995})}\BibitemShut {NoStop}%
\bibitem [{\citenamefont {Akmal}\ \emph {et~al.}(1998)\citenamefont {Akmal},
  \citenamefont {Pandharipande},\ and\ \citenamefont
  {Ravenhall}}]{PhysRevC.58.1804}%
  \BibitemOpen
  \bibfield  {author} {\bibinfo {author} {\bibfnamefont {A.}~\bibnamefont
  {Akmal}}, \bibinfo {author} {\bibfnamefont {V.~R.}\ \bibnamefont
  {Pandharipande}}, \ and\ \bibinfo {author} {\bibfnamefont {D.~G.}\
  \bibnamefont {Ravenhall}},\ }\href {\doibase 10.1103/PhysRevC.58.1804}
  {\bibfield  {journal} {\bibinfo  {journal} {Phys. Rev. C}\ }\textbf {\bibinfo
  {volume} {58}},\ \bibinfo {pages} {1804} (\bibinfo {year}
  {1998})}\BibitemShut {NoStop}%
\bibitem [{\citenamefont {Zhao}\ \emph {et~al.}(2015)\citenamefont {Zhao},
  \citenamefont {Xu}, \citenamefont {Yan}, \citenamefont {Luo}, \citenamefont
  {Liu},\ and\ \citenamefont {Zong}}]{PhysRevD.92.054012}%
  \BibitemOpen
  \bibfield  {author} {\bibinfo {author} {\bibfnamefont {T.}~\bibnamefont
  {Zhao}}, \bibinfo {author} {\bibfnamefont {S.-S.}\ \bibnamefont {Xu}},
  \bibinfo {author} {\bibfnamefont {Y.}~\bibnamefont {Yan}}, \bibinfo {author}
  {\bibfnamefont {X.-L.}\ \bibnamefont {Luo}}, \bibinfo {author} {\bibfnamefont
  {X.-J.}\ \bibnamefont {Liu}}, \ and\ \bibinfo {author} {\bibfnamefont
  {H.-S.}\ \bibnamefont {Zong}},\ }\href {\doibase 10.1103/PhysRevD.92.054012}
  {\bibfield  {journal} {\bibinfo  {journal} {Phys. Rev. D}\ }\textbf {\bibinfo
  {volume} {92}},\ \bibinfo {pages} {054012} (\bibinfo {year}
  {2015})}\BibitemShut {NoStop}%
\bibitem [{\citenamefont {Masuda}\ \emph
  {et~al.}(2013{\natexlab{c}})\citenamefont {Masuda}, \citenamefont {Hatsuda},\
  and\ \citenamefont {Takatsuka}}]{Masuda01072013}%
  \BibitemOpen
  \bibfield  {author} {\bibinfo {author} {\bibfnamefont {K.}~\bibnamefont
  {Masuda}}, \bibinfo {author} {\bibfnamefont {T.}~\bibnamefont {Hatsuda}}, \
  and\ \bibinfo {author} {\bibfnamefont {T.}~\bibnamefont {Takatsuka}},\ }\href
  {\doibase 10.1093/ptep/ptt045} {\bibfield  {journal} {\bibinfo  {journal}
  {Prog. Theor. Exp. Phys.}\ }\textbf {\bibinfo {volume} {2013}} (\bibinfo
  {year} {2013}{\natexlab{c}}),\ 10.1093/ptep/ptt045}\BibitemShut {NoStop}%
\end{thebibliography}%
\end{document}